\documentclass{procDDs}
\usepackage[cp1251]{inputenc}
\usepackage[T2A]{fontenc}
\usepackage{cite}
\title{New representations for square-integrable spheroidal functions}

\author{{V. N. Kovalenko}, \textbf{A. M. Puchkov}}% List of authors with the
{Saint-Petersburg State University, Russia}                 % same affiliation,
{a.puchkov@spbu.ru}
\begin {document}

\maketitle

\index{Kovalenko, V.N.}                              % write this for each author
\index{Puchkov, A.M.}                               % to generate the index

\begin{abstract}
We discuss the solution of boundary value problems that arise after the separation of variables in the Schr\"odinger equation in oblate spheroidal coordinates. The specificity of these boundary value problems is that the singular points of the differential equation are outside the region in which the eigenfunctions are considered. This prevents the construction of eigenfunctions as a convergent series.
To solve this problem, we generalize and apply the Jaffe transformation.
We find the solution of the problem as trigonometric and power series in the particular case when the charge parameter is zero.
Application of the obtained results to the spectral problem for the model of a quantum ring in the form of a potential well of a spheroidal shape is discussed with introducing a potential well of a finite depth.
\end{abstract}

%%% Introduction
\section{Formulation of the problem}

In works \cite{putchkov1,putchkov2,putchkov3,putchkov4,putchkov5} various quantum potential models  that allow separation of variables in the Schr\"odinger equation
\begin{equation}
\label{a002}
\Delta \Psi + 2\left(E - V \right)\Psi = 0
\end{equation}
in spheroidal coordinates were considered. Particular attention was drawn to the problems with separation of variables in oblate spheroidal coordinates due to the specificity of the emerging boundary value problems.
The oblate spheroidal coordinates $(\xi, \eta, \varphi)$
are related to Cartesian ones in the following way:
\begin{gather} \nonumber
\label{a3}\hspace{-0.4cm}
x=\frac{R}{2}\sqrt{(\xi^2+1)(1-\eta^2)}\cos{\varphi}\,,\\
y=\frac{R}{2}\sqrt{(\xi^2+1)(1-\eta^2)}\sin{\varphi}\,,\\
z=\frac{R}{2}\xi\eta\,. \nonumber
\end{gather}
The intervals, where these variables are defined, are traditionally chosen in two different ways \cite{komarov}:
\begin{equation}
\label{a4}
a)\;\;\;  \xi \in [0,\infty),\;\;\; \eta \in
[-1,1],\;\;\; \varphi \in [0,2\pi);
\end{equation}
\begin{equation}
\label{a5}
b)\;\;\; \xi \in (-\infty, \infty),\;\;\; \eta \in
[0,1],\;\;\; \varphi \in [0,2\pi)\,.
\end{equation}

In this paper, we discuss the solution of the homogeneous boundary value problem:
\begin{gather}
\label{b1}
\dfrac{d}{d\xi}(\xi^{2}+1)\dfrac{d}{d\xi}X_{mk}(\xi) - \\ \nonumber
- \left[ \lambda + p^{2}(\xi^{2}+1) - a\xi -
\dfrac{m^{2}}{\xi^{2}+1} \right]X_{mk}(\xi) = 0\,,
\end{gather}
\begin{equation}
\label{b2}
| X_{m k}(\xi)  | \xrightarrow[\xi \to \pm \infty]{}
0, \quad \xi \in (-\infty, \infty)\,,
\end{equation}
where, the values of $k$ and $m$ are integer parameters, and $k$ is the number of zeros of the eigenfunction $X_{m k}(\xi)$
on the interval $(-\infty,+\infty).$

This problem first arose in work \cite{putchkov1} in connection with the
 study of the discrete spectrum of the generalized quantum mechanical problem of two centers; therefore, a square integrability was additionally required:
$X_{mk}(\xi) \in
{\cal L}_{2}\left(\mathbb{R}\right)\,.$

The equation (\ref{b1}) is a Coulomb spheroidal equation on the imaginary axis
$z$: \  $  \xi = \text{Im}\ z.$
It has three singular points:
$\xi_{1} = +\imath,\  \xi_{2} =
-\imath$ and $\xi_{3} = \infty,$
where $\xi_{1}$ and $\xi_{2}$ are regular, and $\xi_{3}$ is irregular.

In the boundary value problem (\ref{b1}) -- (\ref{b2}), due to the configuration of singular points and the domain of an eigenfunction,
one faces a problem of the disk of convergence and, therefore, it is not possible to use the Jaffe transforms \cite{komarov}
for the representation of  $X_{m k}(\xi)$
as a series. Below, we present a generalized Jaffe transformation, which includes the standard one as a particular case.

\section{The generalized Jaffe transformation}

We represent the eigenfunction in the following form:
$$
X_{mk}(\xi) =
\dfrac{e^{-p\sqrt{\xi^2+1}}}
{\sqrt{\xi^2+1}}  F_{mk}(\xi)\,,
$$
where $F_{mk}(\xi)$ asymptotically behaves as
$$
F_{mk}(\xi)
\sim
(\xi + \sqrt{\xi^2 + 1})^{a/2p}
\;\;\; \text{at} \;\;\;
\xi \to +\infty\,,
$$
$$F_{mk}(\xi)
\sim
(\xi - \sqrt{\xi^2 + 1})^{a/2p}
\;\;\; \text{at} \;\;\;
\xi \to -\infty\,.
$$
The basic idea of constructing a generalized Jaffe transformation is to map the singular points (\ref{b1}) to the unit circle
$$
-\imath \mapsto -\imath\,,\;\;\;
+\imath \mapsto +\imath\,,\;\;\;
-\infty \mapsto -1\,,\;\;\;
 +\infty \mapsto +1\,.
$$
by using the substitution
\begin{equation}
\label{ourmap}
t = \dfrac{\xi}{\sqrt{\xi^2 + 1} + 1}\,.
\end{equation}

%${\xi \mapsto x \mapsto t,}$ where
%\begin{gather}\label{ourmap}{\xi = \sinh{x},} \; {t = \tanh{(x/2)}},\end{gather}
%one maps the singular points (\ref{b1}) to the unit circle:
% The basic idea of the constructing a generalized Jaffe transformation is that
% состоит в том, чтобы с помощью подстановки () перевести особые точки
% (\ref{b1}) to the unit circle

\begin{figure}[hbt]
\begin{center}\vspace{-0.3cm}
\includegraphics[width=0.36\textwidth]{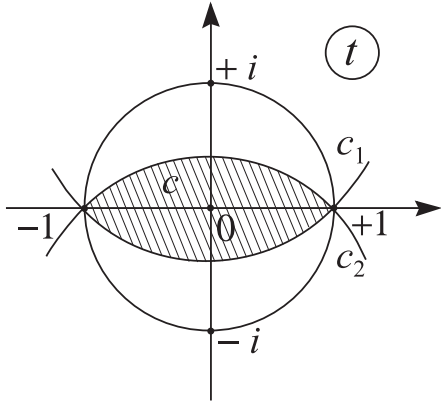}
\caption{Configuration of the range of the variable $t$ and singular points for equation~(\ref{b1}).}
\label{Fig.4b}
\end{center}\vspace{-0.3cm}
\end{figure}

Note that an additional singular point appears at infinity  (Fig.~\ref{Fig.4b}).

The boundary value problem (\ref{b1})--(\ref{b2}) is transformed into the following:

\begin{gather}\nonumber
\dfrac{1}{4}(1-t^2)^2\dfrac{d^2}{dt^2}F_{mk}(t) + \\\nonumber \left[ -2pt -
\dfrac{t(1 - t^2)}{2} - \dfrac{t(1-t^2)}{(1+t^2)} \right]
\dfrac{d}{dt}F_{mk}(t)+ \\ \label{b7a}
+ \left[ -\lambda - p^2 -p\dfrac{(1-t^2)}{(1+t^2)} +
\dfrac{2at}{(1-t^2)} + \right. \\\nonumber
\left. +(m^2 - 1)\dfrac{(1-t^2)^2}{(1+t^2)^2}
\right] F_{mk}(t) = 0\,,
\end{gather}
\begin{gather}
| F_{mk}( \pm 1) | < \infty , \;\;\; -1 \leqslant t \leqslant
+1 \,. \label{b8}
\end{gather}

As it is known \cite{komarov,slavyanov}, the boundary value problem for the Coulomb spheroidal equation
on the interval $[1, \infty)$  appears in the quantum problem of two Coulomb centers (problem $Z_{1}eZ_{2}$).
In this problem, the boundary points are singular points of the differential equation. To solve it, we apply
the Jaffe transformation $y=(\xi-1)/(\xi+1)$. It transforms the domain of variable $\xi$ into the interval $[1, \infty) \to [0, 1)$,
so that $1 \to 0, \infty \to 1.$
Then the eigenfunction is represented in the form of a power series in the variable $y$.
In our problem, we use transformation (\ref{ourmap}) to transfer the domain of the variable $(-\infty,\infty) \to (-1,1)$.

Let us solve the boundary value problem (\ref{b7a}) --  (\ref{b8})
separately on the intervals $(-1,0)$ and $(0,+1)$ applying
an analogue of the Jaffe transformation, and then requiring the resulting solutions to match smoothly at zero.
We say that such transformation is the generalized Jaffe transformation. Note that, in order to obtain a real representation of $F_{mk}(t)$,
it is necessary to allow both  basic monomials:
$$
\left(\dfrac{t-\imath}{t+\imath}\right)^{n-m} \;\;\; \text{and} \;\;\;
\left(\dfrac{t+\imath}{t-\imath}\right)^{n-m}\,.
$$
As a result, we look for the eigenfunctions on each of the intervals in the following form:

1) for $t \in [0, 1):$
\begin{gather}
\label{b9}
F_{mk}(t)=A\left( \dfrac{1+t}{1-t} \right)^{a/2p} \times \\ \nonumber \times
\left[
\sum_{n=0}^{\infty}
f_{n}^{(1)}
\left(\dfrac{t-\imath}{t+\imath}\right)^{n-m} +
f_{n}^{(2)}
\left(\dfrac{t+\imath}{t-\imath}\right)^{n-m}
\right]\,,
\end{gather}

2) for $t \in (-1, 0]:$
\begin{gather}
\label{b09}
F_{mk}(t)=A\left( \dfrac{1-t}{1+t} \right)^{a/2p} \times \\ \nonumber \times
\left[
\sum_{n=0}^{\infty}
g_{n}^{(1)}
\left(\dfrac{t-\imath}{t+\imath}\right)^{n-m} +
g_{n}^{(2)}
\left(\dfrac{t+\imath}{t-\imath}\right)^{n-m}
\right]\,.
\end{gather}

Our ansatz, in contrast to the Jaffe transformation \cite{komarov}, contains not one, but four sets of coefficients.
Coefficients in each of the sets $f_{n}$ and $g_{n}$ are related by similar five-term recurrence relations. For example, for $f_{n}$ we have:
%=======================================================================================================================
%Коэффициенты в каждом из наборов $f_n$ и $g_n$ удовлетворяют похожим (я подчеркиваю - похожим, подобным или аналогичным)  пятичленным рекуррентным соотношениям. Слово same надо заменить.
%=======================================================================================================================
\begingroup
\setlength{\thinmuskip}{1mu}
\setlength{\medmuskip}{2mu}
\setlength{\thickmuskip}{3mu}
\begin{gather}
\nonumber
\left( 1 + \dfrac{\alpha_{2}}{n} + \dfrac{\beta_{2}}{n^2} \right)
f_{n+2} + \left( \dfrac{\alpha_{1}}{n} + \dfrac{\beta_{1}}{n^2}
\right) f_{n+1} + \\ \label{recur_rel1}  + \left( 2 + \dfrac{\alpha_{0}}{n} +
\dfrac{\beta_{0}}{n^2} \right)
f_{n}  + \\ \nonumber + \left( \dfrac{\alpha_{-1}}{n} + \dfrac{\beta_{-1}}{n^2}
\right) f_{n-1} + \left( 1 + \dfrac{\alpha_{-2}}{n} +
\dfrac{\beta_{-2}}{n^2} \right) f_{n-2} = 0\,,
\end{gather}
where
\begin{align*}
\alpha_2:=-2(m-1),\,\  \ \ \ \ & \beta_2:=-(m+1),\\
\alpha_1:=-4\left(p+\dfrac{\imath a}{2p}\right), \,\ &
  \beta_1:=4\left(p+\dfrac{\imath a}{2p}\right)\left(m-\dfrac{1}{2}\right),\\
\alpha_0:=-4m,\ \ \ \ \ \ \ \ \ \ \  \,\ & \beta_0:=4\left(\lambda+p^2-\dfrac{a^2}{4p^2}\right)+2,\\
\alpha_{-1}:=4\left(p-\dfrac{\imath a}{2p}\right), \,\ &
  \beta_{-1}:=-4\left(p-\dfrac{\imath a}{2p}\right)\left(m+\dfrac{1}{2}\right),\\
\alpha_{-2}:=-2(m+1), \ \ \ \,& \beta_{-2}:=(m-1).
\end{align*}
\endgroup
%==========================================================================================
%Для коэффициентов из набора $g_{n}$ имеют место такие же формулы, но с заменой $a \to -a.$
%Оба разложения (\ref{b9}) и (\ref{b09}) должны быть гладко сшиты при $t=0.$ Это означает, что
%мы должны приравнять в нуле не только сами функции, но также и их первые производные. 
%Под равенством надо понимать равенство сумм рядов справа и слева, а не равенство соответствующих слагаемых.
%
% были вещественными. Тогда возникает
%дополнительное ограничение и из четырех независимых наборов коэффициентов останется только два.   

For coefficients $g_ {n}$, similar formulas hold,  with the substitution $a \to -a.$
Both expansions (\ref{b9}) and (\ref{b09}) must be smoothly bound for $t = 0.$ This means that not only the functions themselves, but also their first derivatives must be continuous at $t = 0.$
The equality means that the sums of series on the right and on the left hand side must be equal. We do not require equality of the corresponding summands. Moreover, taking into account that the expressions (\ref{b9}) and (\ref{b09})
should be real, an additional constraint arises that only two of the four independent sets of coefficients remain.
%==========================================================================================

The convergence of the series in (\ref{b9}) is determined by the position of the roots of the characteristic equation:
\begin{gather}
\nonumber
\left( 1 + \dfrac{\alpha_{2}}{n} \right)l^4(n) +
\dfrac{\alpha_{1}}{n}l^3(n) + \left( 2 + \dfrac{\alpha_{0}}{n}
\right)l^2(n) +\\+ \dfrac{\alpha_{-1}}{n}l(n) +
\left(
1 + \dfrac{\alpha_{-2}}{n}
\right)
= 0\,.\label{root1}
\end{gather}

It can be shown that at $n > n_{0},$ ${n_0 = \sqrt{16 a^2/p^2 + 64m^2}}$ two roots (\ref{root1}) lie inside the unit circle. They correspond to the correct sets of coefficients corresponding to the desired eigenfunction. Two more roots are located outside the unit circle. They correspond to a dominant solution (growing at $t \rightarrow \pm 1$), which must be excluded from consideration.

In the series (\ref{b9}) the first term converges uniformly inside the circle $C_{1},$ and the second one -- inside $C_{2}$ (Fig.~\ref{Fig.4b}).
Both series can converge uniformly inside the domain
$C = C_{1}\cap C_{2}$ and, hence, on the interval $(-1, +1),$ which includes $[0, +1).$ \newline
The series (\ref{b09}) converges similarly.

The implementation of the numerical procedure for calculating eigenvalues and eigenfunctions
is similar to that described in the monograph \cite{slavyanov}.
Let us briefly explain it. Recurrent relation (\ref{recur_rel1}) for eigenvalues $\lambda$
and two sets of coefficients can be represented in matrix notation, as follows:
\begin{equation}
\label{b13}
\mathbf{A} \mathbf{f^{(1)}} = 0\,,
\;\;\;\;\;\;\;\;
\mathbf{\tilde{A}} \mathbf{f^{(2)}} = 0\,.
\end{equation}
Since for both sets considerations are very similar, we would work only with the first one.
The precise condition for eigenvalues reads
%Поскольку все рассуждения для обоих наборов совершенно аналогичны, то
%для краткости будем описывать случай только с первым набором.
%Точное условие для определения собственных значений имеет вид:
$$
\det \mathbf{A} = 0\,.
$$
In practice, an infinite matrix $\mathbf{A}$ is truncated at some large but finite $n = N.$
Then, to get rid off Birkhoff's exponentially growing solutions, we need to demand the following conditions:
%На практике, бесконечная матрица $\mathbf{A}$ обрывается при некотором достаточно
%большом $n = N.$ Тогда, чтобы избавиться от экспоненциально растущих решений Биркгофа
%надо потребовать, чтобы выполнялись условия:
$$
f^{(1)}_{N+1} = 0\,, \;\;\;\;
f^{(1)}_{N+2} = 0\,.
$$
Two recessive solutions for the set of $\mathbf{f^{(1)}}$ are obtained from conditions:
%Два рецессивных решения для набора $\mathbf{f^{(1)}}$ получаются из условий:
\begin{equation}
\label{b014}
\tilde{f}^{(1)}_{N} = 1\,,\;\;\; \tilde{f}^{(1)}_{N-1} = 0\,, \;\;\;
\tilde{\tilde{f}}^{(1)}_{N} = 0\,,\;\;\; \tilde{\tilde{f}}^{(1)}_{N-1} = 1\,.
$$
The recessive general solution would be
%Общее рецессивное решение будет иметь вид:
$$
f^{(1)}_{n} =
K_{1} \tilde{f}^{(1)}_{n} +
K_{2} \tilde{\tilde{f}}^{(1)}_{n}
\end{equation}
with arbitrary constants $K_{1}$ and $K_{2}.$
Then we use the reverse recursion, and the convergence condition for it is as follows:
%%с произвольными константами $K_{1}$ и $K_{2}.$
%Далее, используется обратный ход рекурсии, а условие ее обрывания
%сводится к равенству:
  \begin{eqnarray}
 \label{b14}
         \det
         \left(\begin{array}{ll}
         \tilde{f}^{(1)}_{-1} \;\;\;\;\; \tilde{\tilde{f}}^{(1)}_{-1} \\
         \tilde{f}^{(1)}_{-2} \;\;\;\;\; \tilde{\tilde{f}}^{(1)}_{-2}
         \end{array}
         \right) = 0\,.
 \end{eqnarray}
 Coefficient $f_{1}$ can be found from the condition (\ref{b8}), while $f_{0}$ remains arbitrary.
% Коэффициент $f_{1}$ определяется из условия (\ref{b8}), а $f_{0}$ остается произвольным.

\begin{figure}[h]\vspace{-0.3cm}
\includegraphics[width=0.95\columnwidth]{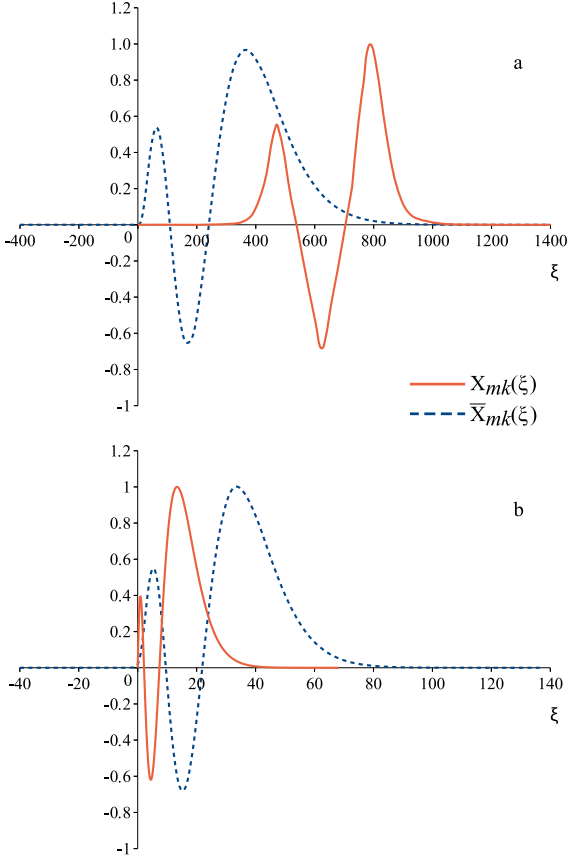}\vspace{-0.01cm}
\caption{
Eigenfunctions on the semi-axis $X_{mk}(\xi)$ (lines) and on the whole axis  $\overline{X}_{mk}(\xi)$ (dashes) for $ k =2, \; m =2.$ a) $a = 0.1\,,$ b) $a = 1\,.$
}\vspace{-0.3cm}
\label{Wavefunc1a}
\end{figure}

The graphs of some eigenfunctions are shown in Fig.~\ref{Wavefunc1a}.

Concluding the section, we should note that the standard Jaffe expansion is a particular case of (\ref{b9}) or (\ref{b09}).\vspace{-0.4cm}

\section{Power expansions of eigenfunctions}

At $a = 0$, the initial-boundary problem (\ref{b1}), (\ref{b2}) acquires additional symmetry related to parity. The equation (\ref{b1}) becomes spheroidal with even and odd eigenfunctions.
We expand them into power series. For even functions, one has:
\begin{equation}
\label{b13a}
F_{mk}^{(1)}(t) = \sum \limits_{n=0}^{\infty}
{f_{1\,{n}\,} \dfrac{(1+t^2)^{n}}{2^{n}}}\,,
\end{equation}
and, for odd ones,
\begin{equation}
\label{b14a}
F_{mk}^{(2)}(t) = t\sum \limits_{n=0}^{\infty}
{f_{2\,{n}\,} \dfrac{(1+t^2)^{n}}{2^{n}}}\,.
\end{equation}

The coefficients of both sets $f_{j\,{n}\,}$ ($j=1,2)$ are related to each other by four-term recurrence relations:
\begin{gather}
\nonumber
\left( -1 + \dfrac{\alpha_{j{2}}}{n} + \dfrac{\beta_{j{2}}}{n^2}
\right) f_{j\,{n+2}} + \\ \nonumber + \left( 4 + \dfrac{\alpha_{j\,{1}}}{n} +
\dfrac{\beta_{j\,{1}}}{n^2} \right) f_{j\,{n+1}} + \\ \label{b15} + \left( -5 +
\dfrac{\alpha_{j\,{0}}}{n} + \dfrac{\beta_{\,j{0}}}{n^2} \right)
f_{j\,{n}} + \\ \nonumber
+ \left( 2 + \dfrac{\alpha_{j\,{\mathrm{-}1}}}{n} +
\dfrac{\beta_{j\,{\mathrm{-}1}}}{n^2} \right) f_{j\,{n-1}} = 0\,.
\end{gather}

For the even set $\{ f_{1\,{n}}\}_{n=0}^{\infty}$ we have the following parameters:
\begin{align*}
\alpha_{1\,2}&:=-2, & \beta_{1\,2}&:=m^{2}-1,\\
\alpha_{1\,1}&:=2p+3, &
\beta_{1\,1}&:=p-2m^{2}+1,\\
\alpha_{1\,0}&:=-4p+2, & \beta_{1\,0}&:=-\lambda-p^2+p+m^{2}-1,\\
\alpha_{1\,-1}&:=-3, &
\beta_{1\,-1}&:=1,
\end{align*}

and for odd one $\{ f_{2\,{n}}\}_{n=0}^{\infty}$:
\begin{align*}
\alpha_{2\,2}&:=-2, & \beta_{2\,2}&:=m^{2}-1,\\
\alpha_{2\,1}&:=2p+5, &
\beta_{2\,1}&:=p-2m^{2}+2,\\
\alpha_{2\,0}&:=-4p-2, & \beta_{2\,0}&:=-\lambda-p^2-p+m^{2}-1,\\
\alpha_{2\,-1}&:=-1, &
\beta_{2\,-1}&:=0.
\end{align*}

The characteristic equation for the recurrence relations (\ref{b15}) in the first order in $n^{-1}$ is
\begin{gather}\nonumber
P_j(l) = \left( -1 + \dfrac{\alpha_{j_{2}}}{n} \right)l^3(n) +
\left( 4 + \dfrac{\alpha_{j_{1}}}{n} \right)l^2(n) +\\+ \left( -5 +
\dfrac{\alpha_{j_{0}}}{n} \right)l(n) + \left( 2 +
\dfrac{\alpha_{j_{-1}}}{n} \right) = 0\,.
\end{gather}
All its roots are real and positive, with one $l_{j\,{1}\,} < 1,$ and the rest two lying outside the unit interval.
Correspondingly, there are three sets of coefficients for the expansion of eigenfunctions. We take the set corresponding to $l_{j\,{1}\,}<1.$
% Процедура вычисления собственных функций аналогична описанной в предыдущем разделе. Более того, она проще, так как необходимо определить всего
% один набор коэффициентов. Заметим, что
% коэффициент $f_{j_{1}}$ также определяется из (\ref{b8}), а $f_{j_{0}}$ остается произвольным.
Determination of eigenfunctions goes analogously. Moreover, it is simpler, because we need to find only one set of coefficients.
 Note that $f_{j\,{1}\,}$ is fixed by (\ref{b8}), while $f_{j\,{0}\,}$ remains arbitrary.

 \vspace{0.1cm}
\section{Application of the results to the model of a quantum ring}
\begin{figure}[h]
%\hspace{-0.33cm}
\includegraphics[width=1.08\columnwidth]{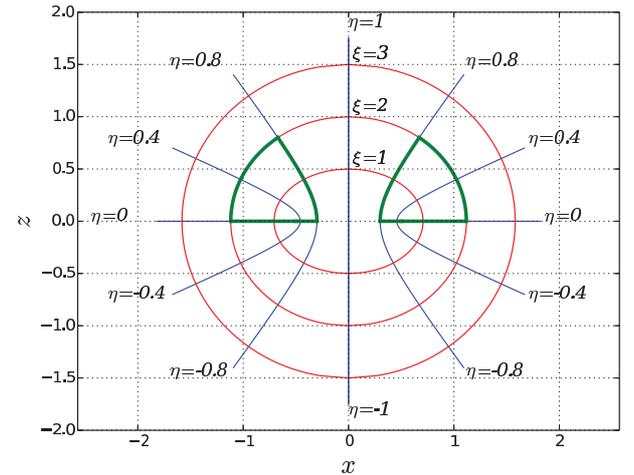}
\caption{
Surfaces $\xi = \text{const}, \; \eta=\text{const},$
projected on the plane $(x,z)$.
The bold line indicates the boundary of the quantum ring. There is a rotational symmetry against z-axis.
}
\label{OSC}
\end{figure}

\vspace{0.1cm}
Let us consider the model of a quantum ring of a spheroidal form with a potential well of finite depth $U_{0}$ (Fig.~\ref{OSC}).
\newline
After separating the variables in the Schr\"odinger equation, we obtain two related boundary-value problems for a quasi-radial equation:

\begin{gather}\label{ring0001}
\dfrac{d}{d\xi}(\xi^{2}+1)\dfrac{d}{d\xi}X^{(1)}_{mk}(\xi) - \\ \nonumber -
\left[\lambda-p^{2}(\xi^{2}+1) -
U_{0}\xi^{2} - \dfrac{m^{2}}{\xi^{2}+1}
\right]X^{(1)}_{mk}(\xi) = 0\,, \\ \nonumber \hspace{-0.11cm}
|X^{(1)}_{mk}(0)| = C_{1}\,, \;\;
|X^{(1)}_{mk}(\xi_{0})| = C_{2}\,, \;\;
0 \leqslant \xi \leqslant \xi_{0}\,,
\end{gather}

\begin{gather}
\label{ring0002}
\dfrac{d}{d\xi}(\xi^{2}+1)\dfrac{d}{d\xi}X^{(2)}_{mk}(\xi) - \\ \nonumber -
\left[\lambda-p^{2}(\xi^{2}+1) -
\dfrac{m^{2}}{\xi^{2}+1}
\right]X^{(2)}_{mk}(\xi) = 0\,, \\ \nonumber \hspace{-0.25cm}
|X^{(2)}_{mk}(\xi_{0})| = C_{2}\,, \;
| X^{(2)}_{m k}(\xi) | \xrightarrow
[\xi \to \infty]{} 0\,, \; \xi_{0} \leqslant \xi < \infty,
\end{gather}
where $C_{1}$ and $C_{2}$ are constants, and
$p^{2} = - \dfrac{|E| R^{2}}{2}$ is an energy parameter.
At the border $\xi = \xi_{0}$, the eigenfunctions must be smoothly bound, so the following conditions must be satisfied:
\begin{gather} \nonumber
X^{(1)}_{mk}(\xi_{0}) =
X^{(2)}_{mk}(\xi_{0}) = C_2\,, \\ \nonumber
\left.
X'^{(1)}_{mk}(\xi)
\right|_{\xi = \xi_{0}} =
\left.
X'^{(2)}_{mk}(\xi)
\right|_{\xi = \xi_{0}}
\,.
\end{gather}

For the solution of the boundary problems (\ref{ring0001}) -- (\ref{ring0002}) one can change the variable as in (\ref{ourmap}) and then use the expansion (\ref{b9}) at $a=0$, or  (\ref{b13a})--(\ref{b14a}) in a general case.

\section{Conclusion}

We proposed a new representation for square integrable spheroidal functions as trigonometrical and power series.
For the solution of this problem, particularly, the generalized Jaffe transform has been applied.
This representation can also be used for the investigation of the spectra of spheroidal quantum rings and quantum dots.

A particular interest may be in applying the asymptotic methods of the type \cite{braun} for the approximation of the functions by ones of a simpler form.

\section{Acknowledgements}
The authors are grateful to the Organizing Committee of the conference Days on Diffraction 2017, especially to Prof. O.~V.~Motygin for numerous discussions and useful comments.

The authors acknowledge Saint-Petersburg State University for the research grant
11.38.242.2015.
%
%==============================================================================
%Авторы благодарны оргкомитету конференции и особенно О. В. Мотыгину (O. V. Motygin) 
%за многочисленные обсуждения и полезные замечания.
%==============================================================================
%%% References
%% Note: use of BibTeX als works!!
\bibliographystyle{unsrt}
%\begin{thebibliography}{1}

%\section{Introduction}
%
%Body of the paper divided into sections and subsections.
%
%\subsection{Subsection}
%
%Body of subsection.
%
%Numbered formula should be written as
%\begin{equation}\label{key} % key is any name used
%a+b=2. % to refer to the formula
%\end{equation}
%
%The reference to such an equation should be as (\ref{key}).
%
%A few formulae:
%\begin{gather} % key is any name used
%a+b=2,\label{key1}\\
%c+d=3.\label{key2}
%\end{gather}
%
%
%Aligned formulae:
%\begin{align} % key is any name used
%A={}&\int_0^1f(x)\mathrm{d}x,\notag\\
%&{}+a+b+c,\label{key3}\\
%B={}&3.\label{key4}
%\end{align}
%
%A long formula:
%\begin{multline}
%D=a+b+c+e+\sum_{n=1}^\infty t_n\\
%+f+\int_0^1g(x)\mathrm{d}x+q+h+y^2\\
%-r-s-\sin(w).
%\end{multline}
%
%% See http://mirror.ctan.org/macros/latex/required/amslatex/math/amsldoc.pdf
%% for more examples of AMS-LaTeX commands.
%
%Citations are written as \cite{paper1}. % paper1 is the name from
% % \bibitem commands below
%Fig.~\ref{fig1} shows handling of figures.
%
%\begin{figure}[t!]\centering
% \includegraphics[width=.43\textwidth]{Test_fig.eps}
% \caption{This is test figure}\label{fig1}
%\end{figure}
%
%\section*{Acknowledgements}
%
%Put acknowledgements in the last section, please do not use footnotes for that.
%
%% citations should be arranged by order of appearance
%
%\begin {thebibliography}{9}
%
%\bibitem{paper1} Surname, I.\,I., Surname, I.\,I., 1998,
% Title of the reference,
% \emph{Journal}, Vol.\;{\bf 2}, pp.\;15--25.
%
%\bibitem{book2} Surname, I.\,I., 2010, \textit{Title of Book}, Publisher, City.
%
%\end{thebibliography}

\end {document}